# Robust topological insulator surface state in MBE grown $(Bi_{1-x}Sb_x)_2Se_3$

Y. Hung Liu[1,‡], C. Wei Chong[1,‡,*], W. Chuan Chen[2], J. C. A. Huang[1,3,4,*], C.-Maw Cheng[2], K.-Ding Tsuei[2], Z. Li[5], H. Qiu[5], V.V. Marchenkov[6]

[1]Department of Physics, National Cheng Kung University, Tainan 70101, Taiwan.

[2]National Synchrotron Radiation Research Center, Hsinchu 300, Taiwan

[3]Advanced Optoelectronic Technology Center (AOTC), National Cheng Kung University Tainan 70101, Taiwan.

[4]Taiwan Consortium of Emergent Crystalline Materials (TCECM), Ministry of Science and Technology, Taipei 10622, Taiwan.

[5]School of Electronic Science and Applied Physics, HeFei University of Technology, Hefei, Anhui 230009, China.

[6]M.N. Miheev Institute of Metal Physics, Ekaterinburg 620137, Russia.

*Correspondence should be addressed to C. Wei Chong (cheongwei2000@yahoo.com) or J. C. Andrew Huang (jcahuang@mail.ncku.edu.tw)

ABSTRACT.

$(Bi_{1-x}Sb_x)_2Se_3$ thin films have been prepared using molecular beam epitaxy (MBE). We demonstrate the angle-resolved photoemission spectroscopy (ARPES) and transport evidence for the existence of strong and robust topological surface states in this ternary system. Large tunability in transport properties by varying the Sb doping level has also been observed, where insulating phase could be achieved at x~0.5. Our results reveal the potential of this system for the study of tunable topological insulator and metal-insulator transition based device physics.

**INTRODUCTION**

During the past decade, three dimensional topological insulators (TIs) have been studied extensively in both the fundamental and technological aspect. Their metallic surface states originated from strong spin-orbit coupling (SOC) and band inversion that are topological protected by time reversal symmetry.[1,2,3,4,5] One of the remarkable feature is the spin-momentum locked (SML) Dirac-like surface band that enables electrically controlled spin polarization in TIs channel.[6,7,8,9,10] These unique surface states make them the promising candidates for future high speed/low power electronics and spintronic devices. Nevertheless, utilization of the SML for device applications remains a challenge since the surface carrier conductions are always overwhelmed by the contribution from the bulk carriers which are non-topological protected.[11,12,13,14] Unique method for extraction of the surface contribution is a necessary step for the development of TI-based device applications.

Utilizing the band structure engineering, $(Bi_{1-x}Sb_x)_2Te_3$ ternary compound has been proven a successful approach in achieving the ideal TI with truly insulating bulk.[15] With reducing the bulk carrier density by over two orders of magnitude, a clear ambipolar gating effect in $(Bi_xSb_{1-x})_2Te_3$ nanoplate had been demonstrated[16] However, no topological phase transition could be observed in this system, where the topological surface states (TSS) are shown to exist over the entire composition range of $(Bi_{1-x}Sb_x)_2Te_3$.[15] Recent attention has been paid to the issues regarding the phase transition between TI and non-topological metal or band insulator (BI) owing to its exotic quantum phenomena and the versatility for device fabrication.[17,18,19,20] To date, detailed studies in both experimental and theoretical works mostly focused on $(Bi_{1-x}In_x)_2Se_3$ in which Brahlek et al. demonstrated the transition from topological metal to a band insulator at x~0.25.[18] Nevertheless, the system became non-topological metal even at low doping level as low as x ~0.03-0.07.[18] It is desired to search for a system where transition between ideal TI and BI can be obtained, where large changes in the surface transport properties could be expected.

$(Bi_{1-x}Sb_x)_2Se_3$ is another emerged candidate that exhibits TI surface states even at large x, where the critical concentration $x_c$ ~0.78-0.83, as predicted theoretically.[21] $Bi_2Se_3$ is a topological insulator with a bulk band gap of 0.3 eV, while $Sb_2Se_3$ is a trivial insulator with a band gap around 1.3 eV. Both the band gap and SOC strength can be modified when $Bi_2Se_3$ is doped with Sb, leading to a topological phase transition.[21,22] Experimentally, Zhang et al. demonstrated the great reduction of bulk carrier density in MBE grown $(Bi_{1-x}Sb_x)_2Se_3$ at x=0.3, where the Dirac-cone like SS showed similar structure and dispersion as undoped $Bi_2Se_3$ as evidenced from angle-resolved photoemission spectroscopy (ARPES).[23] Via transport measurement, Zhang et al. presented the metal-insulator transition (MIT) at x~0.8 that was suggested to be attributed to topological phase transition.[22] Another group, instead of using MBE, Lee et al. reported the MIT in $(Bi_{1-x}Sb_x)_2Se_3$ nanosheet at x~0.22 that prepared by van der Waal epitaxy method.[24] Owing to the importance of this system for the study of TI phase transition, systematic studies on the material growth and characterization, including ARPES and transport evidence for the existence of robust TSS are highly desirable. Here, we fabricated $(Bi_{1-x}Sb_x)_2Se_3$ TI films using molecular beam epitaxy. Enhanced surface states transport was observed via high field Hall effect and weak antilocalization measurement when increasing the Sb doping level. On the other hand, temperature-dependent sheet resistance, $R_s$ demonstrated the metal-insulator transition for the sample with x=0.5, where the $R_s$ (at 300 K) as large as ~18 kΩ about six-time larger than that of x=0.32. The transition occurred around 100 K that is believed due to the competition between the insulating (BI phase) and metallic phase (TI phase). More importantly, ARPES revealed the signature of strong TSS up to x=0.32. Our results reveal the strong potential of this ternary $(Bi_{1-x}Sb_x)_2Se_3$ TI for tunable topological insulator and metal-insulator transition device applications, where wide range of transport properties could be achieved.

**EXPERIMENTAL METHODS**

A series of $(Bi_{1-x}Sb_x)_2Se_3$ thin film were prepared using molecular beam epitaxy (AdNaNo made MBE-9 system). The doping level, x was controlled by fixing the Bi/Se flux ratio, while Sb flux (in Å /min) was changed by varying the source temperature. The c-plane $Al_2O_3$ (0001) substrate was used for the growth of $(Bi_{1-x}Sb_x)_2Se_3$ thin films. High-purity Bi(99.99%), Se(99.999%) and Sb(99.999%) were evaporated using Knudsen cells. All the films used in this work have a thickness of ~25 quintuple-layers (QLs) with Se capping layer ~1 nm that was deposited *in-situ* after the growth of TI layer to avoid environmental contamination.[14] *In situ* reflection high energy electron diffraction (RHEED) was used for monitoring the film quality. Structural characterization was performed by X-ray diffraction (XRD), high-resolution transmission electron microscopic (HRTEM) and Raman spectroscopy, while the surface morphology was obtained by atomic force microscopy (AFM). For the electrical transport measurement, the TI films were patterned into Hall bar geometry using photolithography, allowing the measurement of longitudinal resistance ($R_{xx}$) and Hall resistance ($R_{xy}$). The angle-resolved photoemission spectroscopy (ARPES) experiment was performed at the National Synchrotron Radiation Research Center in Hsinchu, Taiwan using the U9-CGM spectroscopy beamline. The chamber base pressure was ~$10^{-10}$ torr and all the measurements were performed at 65 K. The samples for the ARPES were prepared using the same method as our previous works.[25]

**RESULTS AND DISCUSSION**

The composition of the $(Bi_{1-x}Sb_x)_2Se_3$ thin films were determined using TEM-EDS (energy dispersive X-ray spectroscopy) as shown in Figure S1. Figure 1a-d shows the RHEED patterns for samples with various doping level. Streaky patterns were observed for all the samples, indicating the high-quality single crystalline films were formed. Figure 1e displays the HRTEM image of the $(Bi_{0.68}Sb_{0.32})_2Se_3$ film, clearly demonstrating the epitaxial growth of quintuple layers (QLs) of the TI film. For the composition at x=0.5, the streak pattern becomes blur out, revealing the crystal structure has been disturbed. The evolution of the RHEED pattern is consistent with the XRD data as shown in Figure 1f. All the samples exhibit single phase that crystallized in rhombohedral structure (SG R-3m, Z=3) with (00l) c-axis orientation except for the sample with x=0.5. The extra peaks were identified as (020), (130), (240), (061) and (370), corresponding to the diffraction signals from $Sb_2Se_3$ that crystallized in orthorhombic phase (Figure S2).

$Bi_2Se_3$ exhibits the characteristic triangular terraces and steps as revealed by AFM shown in Figure S3a. When Sb is doped into $Bi_2Se_3$ (Figure S2b and S2c for x=0.28 and 0.32 respectively), triangular terraces on the surface of the $(Bi_{1-x}Sb_x)_2Se_3$ films become smaller with higher surface roughness. Most remarkable feature is observed at the sample $(Bi_{0.5}Sb_{0.5})_2Se_3$ (Figure S3d), where needle-like morphology appeared in which the surface roughness as high as ~4 nm (rms value). As revealed from XRD, it was suggested that the rough morphology originated from the precipitation of $Sb_2Se_3$ phase.

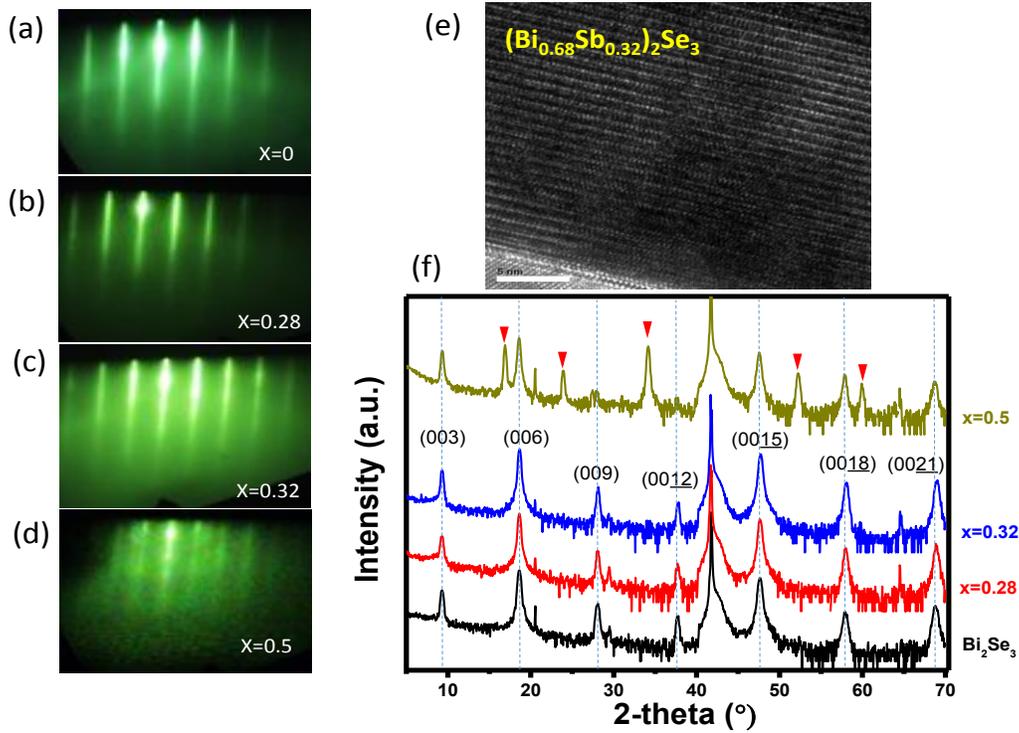

**Figure 1.** (a)-(d) RHEED patterns for $(Bi_{1-x}Sb_x)_2Se_3$ thin films; (e) TEM image for x=0.32 (scale bar: 5 nm); (f) X-ray diffraction data for x=0, 0.28, 0.32, 0.5. The dashed lines indicate the peaks corresponding to the (003n) family, indicating the c-axis orientation growth. The triangular symbols reveal the peaks of orthorhombic $Sb_2Se_3$.

Figure 2a displays the observed Raman spectra for the samples at x=0, 0.28, 0.32 and 0.5. $Bi_2Se_3$ is a strong Raman active material that can be identified by their characteristic phonon modes of $E^1_g$, $A^1_{1g}$, $E^2_g$ and $A^2_{1g}$ in the low wave number region.[25,26] While the accessible range of our instrument is 50-250 cm$^{-1}$, the Raman spectra of the $Bi_2Se_3$ film (x=0) clearly displays $A^1_{1g}$, $E^2_g$ and $A^2_{1g}$ Raman peaks that corresponding to an in-plane ($E^2_g$) and two out of plane ($A^1_{1g}$ and $A^2_{1g}$) vibrational modes of the (-Se(1)-Bi-Se(2)-Bi-Se(1)-) lattice.[25,26] Interestingly, the vibration modes of $E^2_g$ and $A^1_{1g}$ are not changed significantly with increasing the Sb content (Figure 2b), indicating the rhombohedral crystalline structure is largely preserved except for x=0.5. Besides, signature of the substitution of Sb into Bi lattice was observed as the $A^2_{1g}$ mode shifted towards high wave number side (Figure 2b, shown as the dashed line.) Due to the incorporation of Sb, some Bi-Se bonds (~175 cm$^{-1}$) are replaced by Sb-Se bonds (~190 cm$^{-1}$) that resulting in the peak shifting.[22,24] For the sample with x=0.5, peaks broadening has been observed, where there are two broad peaks emerged at ~175 and ~190 cm$^{-1}$. The result indicated the coexistence of the rhombohedral and orthorhombic phases at x=0.5, which agrees well with XRD observations.

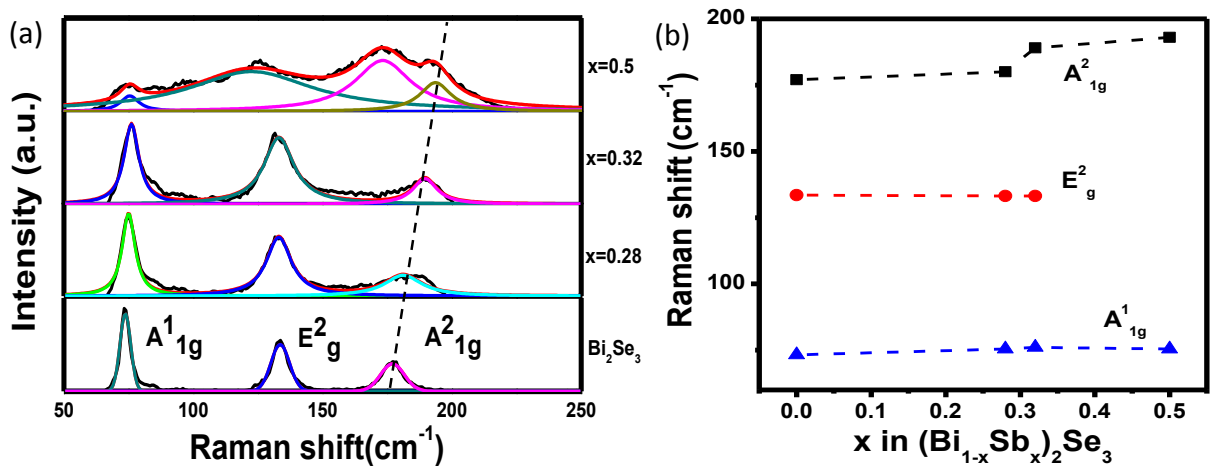

**Figure 2.** (a) Typical Raman spectra of $(Bi_{1-x}Sb_x)_2Se_3$ with different doping levels x=0, 0.28, 0.32, 0.5; (b) Raman shift of $A^1_{1g}$, $A^2_{1g}$ $E^2_g$ vibration modes for various samples.

We further investigate the doping effect by measuring the electrical transport properties. Hall effect measurement, $R_{xy}$ *vs.* B has been presented in Figure 3a in which the nonlinear Hall resistance was observed for the Sb doped samples, indicating the multiple channels contributing to the Hall effect.[18] By taking $dR_{xy}/dB$, the degree of nonlinearity could be revealed as shown in the inset. Very much larger nonlinearity, in comparison to the undoped case, where the Sb doped $Bi_2Se_3$ exhibits steeper slope of $dR_{xy}/dB$ *vs.* B and larger $dR_{xy}/dB$. This finding demonstrates the enhancement of the surface states conduction[18,24,27] when Sb is doped into $Bi_2Se_3$ up to x=032. One of the reason is the depletion/suppression of the contribution of the bulk carrier transport. To obtain the total carrier density, we extract the slope at the high field region as shown in the graph of $R_{xy}$ *vs.* B (dashed line). As expected, Figure 3b shows that the carrier density, $n_{2D}$ decreases by doping the Sb into $Bi_2Se_3$. The total reduction of carrier density, $\Delta n_{2D}$ at doping level of x=0.32 was $\sim 2\times 10^{13}$ cm$^{-2}$ which is almost 70% reduction in comparison with the undoped $Bi_2Se_3$. Due to the same valency of Sb as Bi atoms, no additional charges would be introduced by substitution of Bi with Sb. However, replacement of Bi by Sb could result in the reduction of unit cell volume that enhances the formation energy of Se vacancies. Thus, the decreasing of $n_{2D}$ is attributed to the suppression of Se vacancies in this $(Bi_{1-x}Sb_x)_2Se_3$ ternary compound.[23] On the other hand, sheet resistance $R_s$ increases with increasing the Sb content (Figure 3(c); left axis), following by the degradation of mobility (right axis). For x=0.5, about six-time larger $R_s$ (at 300 K) was observed in comparison to x=0.32 as shown in the $R_s$-T (Figure 3d). In contrast to the metallic behavior (presented by x=0 and x=0.32), the sample (x=0.5) exhibits insulating behavior where $R_s$ increases with decreasing temperature until 100 K. This observation is in agreement with the XRD, AFM and Raman analysis, where coexistence phases of rhombohedra/orthorhombic were found at x=0.5. $Sb_2Se_3$ is a band insulator in nature, the large enhancement of the $R_s$ is believed attributed to the existence of $Sb_2Se_3$ that resulted in the metal-insulator transition as shown in the $R_s$-T.

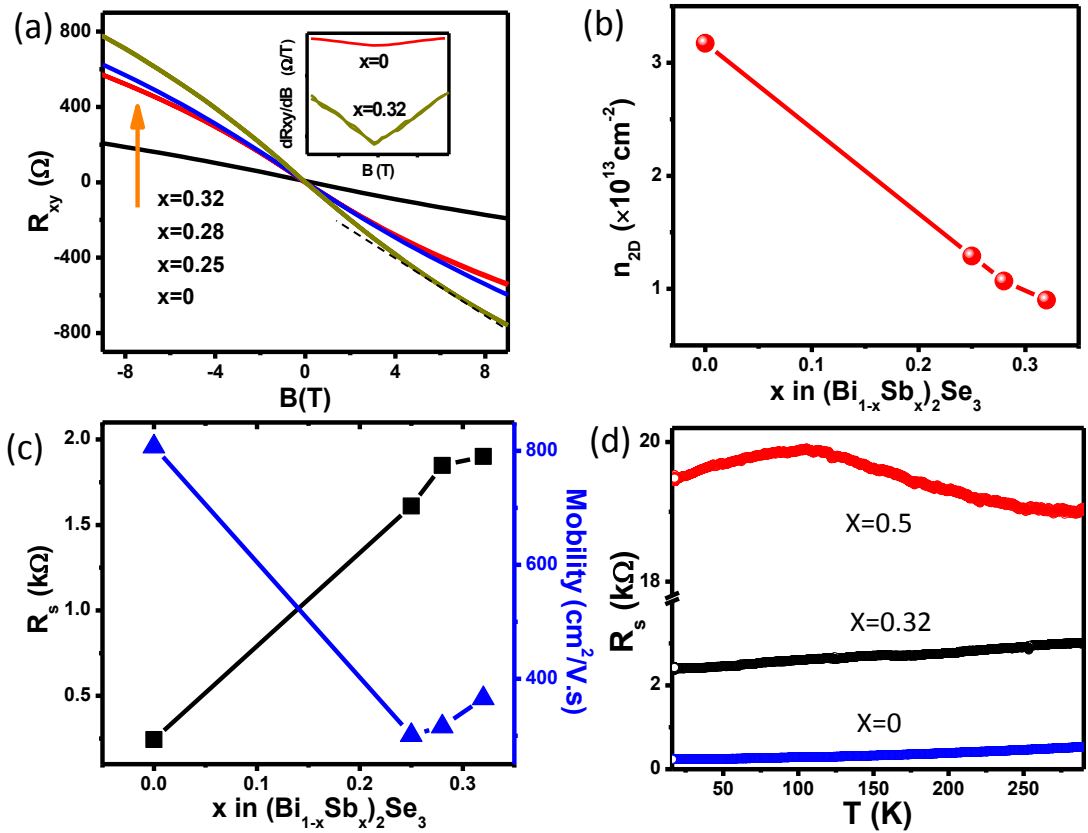

**Figure 3. (a) Hall resistance $R_{xy}$ *vs*. B for x=0, 0.25, 0.28, 0.32; (b) Carrier density $n_{2D}$ for x=0, 0.25, 0.28 and 0.32 that extracted using Hall effect; (c) Sheet resistance $R_s$ (right axis) and Hall mobility for x=0, 0.25, 0.28 and 0.32. All the measurements performed at 1.9 K; (d) $R_s$ vs. T for x=0, 0.32 and 0.5.**

The presence of the nonlinearity in the $R_{xy}$ vs. B indicating the enhancement of the surface states transport for the Sb-doped samples. To further confirm this observation, $MR_{xx}$ vs. B has been measured as shown in Figure 4a. The cusp-like $MR_{xx}$ curves were observed (at low field region), revealing weak antilocalization (WAL) effect that originated from the high SOC of the materials. To obtain quantitative analysis of those measurements, we fit the $MR_{xx}$ with Hikami-Larkin-Nagaoka (HLN) equation that illustrates the 2D behavior of the WAL:[27]

$$\Delta\sigma = \alpha \frac{e^2}{\pi h}\left[\psi\left(\frac{1}{2} + \frac{B_\varphi}{B}\right) - \ln\left(\frac{B_\varphi}{B}\right)\right] \qquad \text{Eq. (1)}$$

Here $\alpha$ is expected to be -1/2 for single coherent channel, $\psi$ is digamma function and $B_\varphi = \hbar/4el_\varphi^2$ is characteristic field, $l_\varphi = (D\tau_\varphi)^{1/2}$ is the phase coherent length, $\tau_\varphi$ is the phase coherent time and D is diffusion constant. All the curves could be well fitted with this HLN equation indicating the presence of TSS transport in these $(Bi_{1-x}Sb_x)_2Se_3$ system.[27] The characteristic coefficient $\alpha$ has been extracted from the fitting where it reveals the number of transporting coherent channels of the 2D system. Consistent with the observation from $R_{xy}$ vs. B, $|\alpha|$ increases with increasing the Sb doping level (from ~0.4 to ~0.6) (Figure 4c), confirming that the reduction of the carrier density in $(Bi_{1-x}Sb_x)_2Se_3$ plays the role in enhancing the transport contribution of topological surface states.

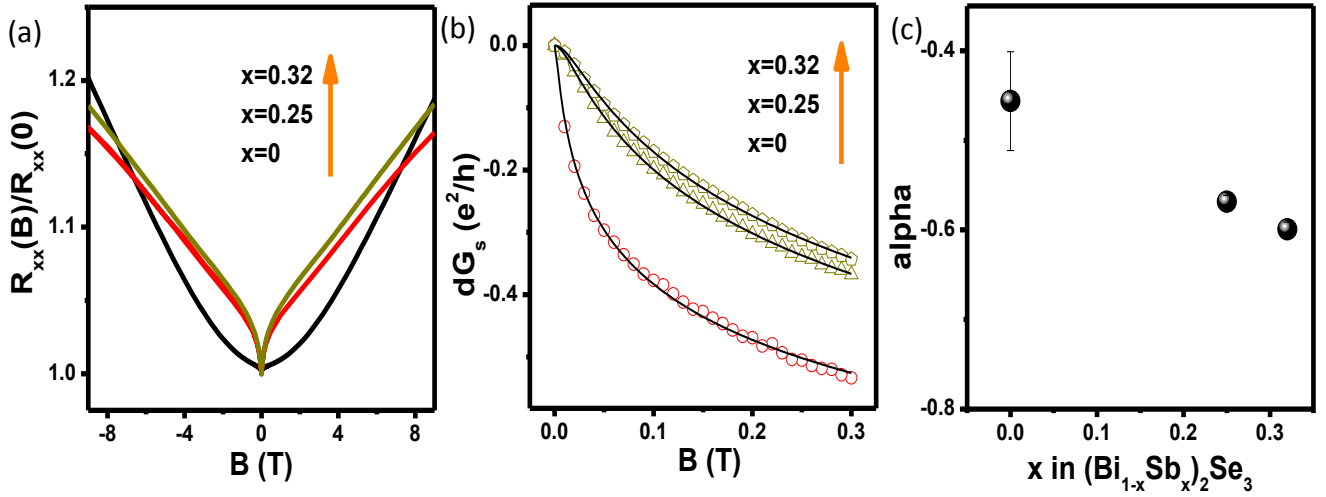

**Figure 4.** (a) $R_{xx}(B)/R_{xx}(0)$ *vs*. B for $(Bi_{1-x}Sb_x)_2Se_3$ with x=0, 0.25 and 0.32; (b) Fitting of $dG_s(e^2/h)$ *vs*. B using HLN equation (solid line indicates fitting curve); (c) alpha extracted from (b) for various samples. All the measurements performed at 1.9 K.

Finally, ARPES measurement has been performed to confirm the existence of the topological surface states in this $(Bi_{1-x}Sb_x)_2Se_3$ system. As shown in Figure 5, linear energy-momentum dispersion was evidenced by ARPES in which strong TSS was observed up to x=0.32. The results reveal that the Fermi level is shifted towards bottom of conduction band (arrows is label at Dirac point for clarity) with increasing the Sb content, indicating the depletion of electron that was in agreement with the Hall measurement result. Interestingly, for the case of x=0.5, where the mixed phase of rhombohedral/orthorhombic occurred, the TSS signal still observable although becomes blur or less obvious. This finding has proven the strong and robust TSS in this $(Bi_{1-x}Sb_x)_2Se_3$ ternary system. At last, we compare the above finding with the theoretical prediction on the topological phase transition in $(Bi_{1-x}Sb_x)_2Se_3$. According to the first-principles calculations, the critical concentration is $x_c$~0.78-0.83 due to the decrease of spin-orbit coupling (SOC) strength.[21] Bulk band gap closing will be happened before the formation of the topological trivial band insulator phase. The prediction was supported by the transport data demonstrated by Zhang et al. where $x_c$~0.8 was determined.[22] However, in our case, the fact could not be concluded due to the high photon energy was used in this experiment, where the precise bulk band gap hardly determined from the ARPES spectrum. Here, we suggest that the enhancement of $R_s$ and metal-insulator transition that observed at x=0.5 should be attributed to mixed phase of rhombohedral/orthorhombic. The evolution of the electronic properties is more appropriately explained based on the scenario, where formation of insulating $Sb_2Se_3$ phase has block the conduction electrons and leading to the MIT.

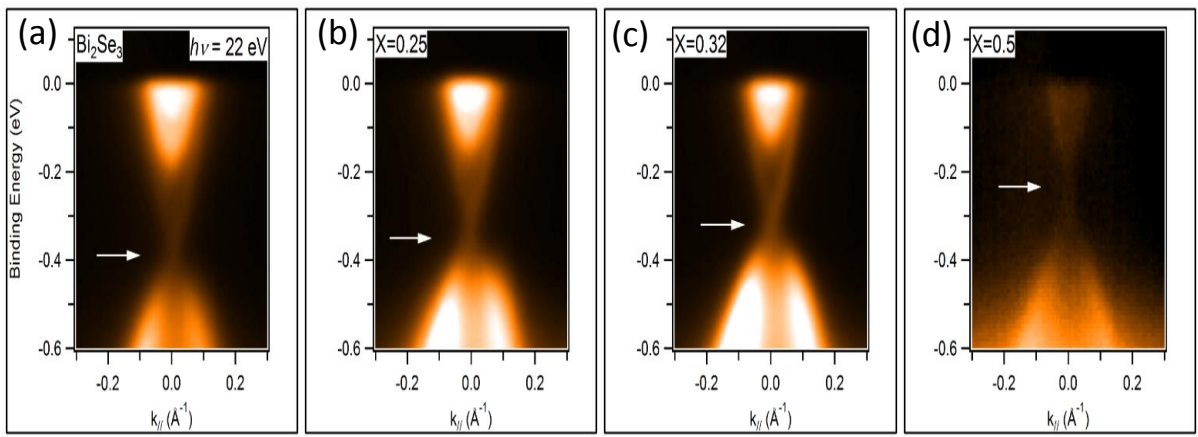

**Figure 5. (a)-(d) ARPES spectra of $(Bi_{1-x}Sb_x)_2Se_3$ for x=0, 0.25, 0.32 and 0.5 that measured at 65 K.**

**CONCLUSION**

In summary, we have synthesized high-quality $(Bi_{1-x}Sb_x)_2Se_3$ thin films using MBE. Large reduction of carrier density $n_{2D}$, where ~70% lower than the undoped $Bi_2Se_3$ was achieved with increasing Sb content. Enhancement of the TSS transport owing to the Sb doping was revealed via the high field Hall effect and weak antilocalization measurement. Large enhancement of sheet resistance and metal-insulator transition (MIT) occurred at x=0.5 that was attributed to the formation of the insulating $Sb_2Se_3$ phase. The topological surface state was further confirmed by ARPES and the linear dispersion was observed up to x=0.5. Our results indicate the robust and strong TSS in this $(Bi_{1-x}Sb_x)_2Se_3$ ternary system where large tunability in transport properties can be achieved, suggesting it could be a promising candidate for exploring the physics and technology of tunable topological insulator and metal-insulator transition.

ASSOCIATED CONTENT

**Supporting Information**. EDS measurement of $(Bi_{1-x}Sb_x)_2Se_3$ grown on sapphire. XRD diffraction patterns for $(Bi_{1-x}Sb_x)_2Se_3$ (BSS) and $Sb_2Se_3$ films that grown using MBE. AFM images of $(Bi_{1-x}Sb_x)_2Se_3$ with x=0, 0.28, 0,32 and 0.5. This material is available free of charge via the Internet at http://pubs.acs.org.

AUTHOR INFORMATION

**Corresponding Author**

(C. Wei Chong) cheongwei2000@yahoo.com; (J. C. Andrew Huang) jcahuang@mail.ncku.edu.tw

**Author Contributions**

The manuscript was written through contributions of all authors. All authors have given approval to the final version of the manuscript. ‡These authors contributed equally.

**ACKNOWLEDGMENTS**

This work was partly supported by the National Science Council (Grant No 104-2811-M-006-029 and 104-2811-M-006-031) and by the Russian Foundation for Basic Research (Grant No 14-02-92012).

(11) Checkelsky, J. G.; Hor, Y. S.; Cava, R. J.; Ong, N. P. Bulk Band Gap and Surface State Conduction Observed in Voltage-Tuned Crystals of the Topological Insulator $Bi_2Se_3$. *Phys. Rev. Lett.* **2009**, 106, 196801.

(12) Butch, N. P.; Kirshenbaum, K.; Syers, P.; Sushkov, A. B.; Jenkins, G. S.; Drew, H. D.; Paglione, J. Strong surface scattering in ultrahigh-mobility $Bi_2Se_3$ topological insulator crystals. *Phys. Rev. B* **2010**, 81, 241301(R).

(13) Taskin, A. A.; Ando, Y. Quantum oscillations in a topological insulator $Bi_{1-x}Sb_x$. *Phys. Rev. B* **2009**, 80, 085303.

(14) Liu, Y. H.; Chong, C. W.; Jheng, J. L.; Huang, S. Y.; Huang, J. C. A.; Li, Z.; Qiu, H.; Huang, S. M.; Marchenkov, V. V. Gate-tunable coherent transport in Se-capped $Bi_2Se_3$ grown on amorphous SiO2/Si. *Appl. Phys. Lett.* **2015**, 107, 012106.

(15) Zhang, J.; Chang, C. Z.; Zhang, Z.; Wen, J.; Feng, X.; Li, K.; Liu, M.; He, K.; Wang, L.; Chen, X.; Xue, Q. K.; Ma, X.; Wang, Y. Band structure engineering in $(Bi_{1-x}Sb_x)_2Te_3$ ternary topological insulators. *Nat. Commun.* **2011**, DOI: 10.1038/ncomms1588.

(16) Kong, D.; Chen, Y.; Cha, J. J.; Zhang, Q.; Analytis, J. G.; Lai, K.; Liu, Z.; Hong, S. S.; Koski, K. J.; Mo, S. K.; Hussain, Z.; Fisher, I. R.; Shen, Z.-X.; Cui, Y. Ambipolar field effect in the ternary topological insulator $(Bi_xSb_{1-x})_2Te_3$ by composition tuning. *Nat. Nanotech.* **2011**, 6, 705-709.

# Supporting Information

## Robust topological insulator surface state in MBE grown $(Bi_{1-x}Sb_x)_2Se_3$

Y. Hung Liu[1,a], C. Wei Chong[1,a,b], W. Chuan Chen[2], J. C. A. Huang[1,3,4,b], C.-M. Cheng[2], K.-D. Tsuei[2], Z. Li[5], H. Qiu[5], V.V. Marchenkov[6]

[1]*Department of Physics, National Cheng Kung University, Tainan 70101, Taiwan.*

[2]*Advanced Optoelectronic Technology Center (AOTC), National Cheng Kung University Tainan 70101, Taiwan.*

[3]*Taiwan Consortium of Emergent Crystalline Materials (TCECM), Ministry of Science and Technology, Taipei 10622, Taiwan.*

[4]*School of Electronic Science and Applied Physics, HeFei University of Technology, Hefei, Anhui 230009, China.*

[5]*M.N. Miheev Institute of Metal Physics, Ekaterinburg 620137, Russia.*

---

[a] Y. Hung Liu, C. Wei Chong contributed equally to this work.

[b] C. Wei Chong or J. C. Andrew Huang: Authors to whom correspondence should be addressed. Electronic mail: cheongwei2000@yahoo.com; jcahuang@mail.ncku.edu.tw

**FIG. S1.** Table below shows the composition of the MBE grown $(Bi_{1-x}Sb_x)_2Se_3$ thin films that determined using TEM-EDS.

| Sb flux Å /min | Component | Sb fraction (%) |
|---|---|---|
| 0.13 | $(Bi_{0.75}Sb_{0.25})_2Se_3$ | 25% |
| 0.3 | $(Bi_{0.72}Sb_{0.28})_2Se_3$ | 28% |
| 0.45 | $(Bi_{0.68}Sb_{0.32})_2Se_3$ | 32% |
| 0.6 | $(Bi_{0.5}Sb_{0.5})_2Se_3$ | 50% |

**FIG. S2.** XRD diffraction patterns for $(Bi_{1-x}Sb_x)_2Se_3$ (BSS) and $Sb_2Se_3$ films that grown using MBE. The extra peaks (pointed by the yellow lines) appeared in the BSS with x=0.5, identified as the peaks corresponding to $Sb_2Se_3$ orthorhombic phase. [1]

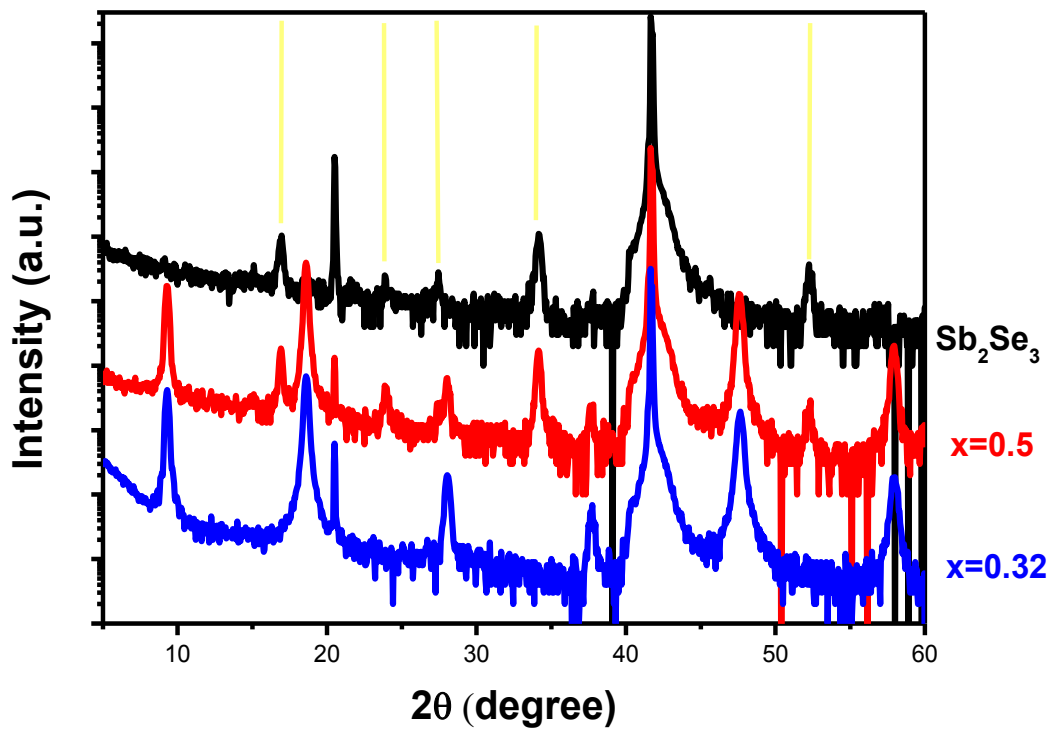

[1] Zhai et al., Adv. Mater. 4530–4533, 22 (2010).

**FIG. S3.** (a)-(d) AFM images of $(Bi_{1-x}Sb_x)_2Se_3$ with x=0, 0.28, 0,32 and 0.5. Scale bar: 500 nm. Table shows the root mean square roughness determined from AFM.

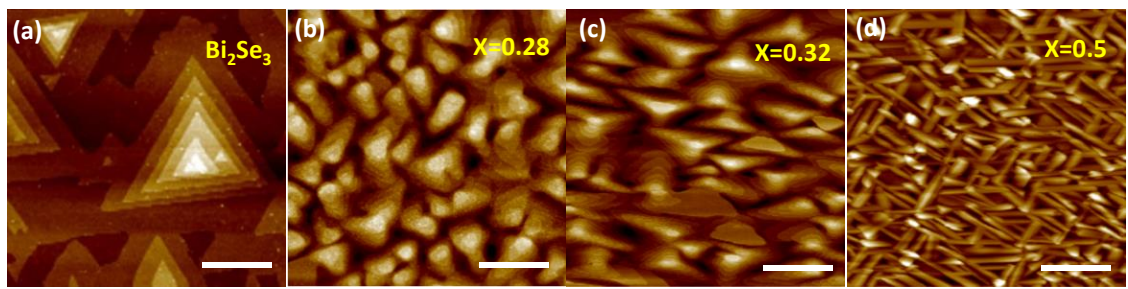

| x in (Bi1-xSbx)2Se3 | $R_q$ (nm) |
|---|---|
| 0 | 1.0 |
| 0.28 | 1.75 |
| 0.32 | 1.93 |
| 0.5 | 4.29 |